\documentclass[aps,twocolumn,showpacs]{revtex4}
\usepackage{graphicx}
\usepackage{amsfonts}
\usepackage{amssymb}
\usepackage{amsbsy}
\usepackage{amsmath}
\usepackage{mathrsfs}
\usepackage{latexsym}
\usepackage{natbib}
\usepackage{bm}
\usepackage{subfigure}
\usepackage{color}
\usepackage{wasysym}

\def\kr{\kappa}

\def\ksg{\mathrm{\varkappa}}

\def\Phit{\tilde{\Phi}}
\def\omegat{\tilde{\omega}}

\def\rs{r_s}
\def\rstar{r_{\star}}

\def\rhash{r_{\sharp}}
\def\scriplus{\mathscr{I}^{+}}
\def\scriminus{\mathscr{I}^{-}}

\def\observerminus{\mathbb{O}^{-}}
\def\observerplus{\mathbb{O}^{+}}

\def\mstar{m_{\star}}

\begin{document}

\title{Consistent derivation of the Hawking effect for both non-extremal
and extremal Kerr black holes}

\author{Subhajit Barman}
\email{sb12ip007@iiserkol.ac.in}

\author{Golam Mortuza Hossain}
\email{ghossain@iiserkol.ac.in}

\affiliation{Department of Physical Sciences, 
Indian Institute of Science Education and Research Kolkata,
Mohanpur - 741 246, WB, India}
 

\date{\today}

\begin{abstract}

It is believed that extremal black holes do not emit Hawking radiation as 
understood by taking extremal limits of non-extremal black holes. However, it is 
debated whether one can make such conclusion reliably starting from an extremal 
black hole, as the associated Bogoliubov coefficients which relate ingoing and 
outgoing field modes do not satisfy the required consistency condition. We 
address this issue in a canonical approach firstly by presenting an exact 
canonical derivation of the Hawking effect for non-extremal Kerr black holes. 
Subsequently, for extremal Kerr black holes we show that the required 
consistency condition is satisfied in the canonical derivation and it produces 
zero number density for Hawking particles. We also point out the reason behind 
the reported failure of Bogoliubov coefficients to satisfy the required 
condition. 

\end{abstract}

\maketitle

\section{Introduction}

In a landmark article \cite{hawking1975}, Hawking pioneered the idea of black 
hole radiation. In particular, by considering quantum fields in static, 
charged or rotating black hole spacetimes, he showed that the asymptotic 
observers would perceive thermal particle creation which is referred to as 
the Hawking effect. In order to derive the Hawking effect, he used ingoing 
and outgoing null coordinates for describing the scalar field modes. In last 
four decades the Hawking effect has been an extensively studied topic of modern 
physics. However, there are some related issues which are still debated, 
particularly involving the case of extremal black holes.

It is usually believed that extremal black holes do not exhibit Hawking 
radiation as one would conclude by taking the extremal limits of non-extremal 
black holes. However, whether one can make such conclusion starting from an 
extremal black hole is still debated in the literature 
\cite{Vanzo:1995bh,Liberati:2000sq, Alvarenga:2003jd, Alvarenga:2003tx}. These 
debates stem from the fact that the associated Bogoliubov transformation 
coefficients that relate the ingoing and the outgoing field modes do not satisfy 
the required consistency relation arising from the commutator brackets between 
the creation and annihilation operators of the field modes. Therefore, these 
Bogoliubov coefficients which are used for computing number density of Hawking 
quanta, are not reliable. Consequently, for extremal black holes it is an 
important question to ask whether one could find a fully consistent derivation 
to conclude about the vanishing Hawking radiation.

In this article, our aims are two fold. Firstly, we show that using 
the so called near-null coordinates which were introduced for computing Hawking 
effect in Schwarzschild spacetime \cite{Barman:2017fzh}, one can perform 
an exact canonical derivation for non-extremal rotating Kerr black holes. The 
usage of these near-null coordinates were necessitated due to the fact that 
null coordinates cannot be used to construct a non-trivial matter field 
Hamiltonian. Consequently, a Hamiltonian based canonical derivation of the 
Hawking effect for a Kerr black hole is still missing.

Secondly, for extremal Kerr black holes, we show that the analogous consistency 
condition which arises from the requirement of the Poisson bracket of field 
modes and their conjugate momenta be simultaneously satisfied for different 
observers, is also fulfilled. Further, we show that in the canonical 
derivation the associated number density operator for the Hawking quanta 
vanishes for the extremal Kerr black holes. This feature reaffirms that the 
extremal Kerr black holes do not emit Hawking radiation.
Additionally, the canonical derivation of the Hawking effect for Kerr black 
holes as presented here provide the initial stage for the study of Hawking 
effect in the context of the so called polymer quantization 
\cite{Ashtekar:2002sn,Halvorson-2004-35}, a canonical quantization method used 
in loop quantum gravity\cite{Ashtekar:2004eh, Rovelli2004quantum, 
Thiemann2007modern}. It appears that the existence of a new length scale could 
substantially affect the Unruh effect \cite{Hossain:2014fma,Hossain:2016klt, 
Hossain:2015xqa} as well as Hawking effect in Schwarzschild spacetime 
\cite{Barman:2017vqx}. However, the question remains whether such claims can be 
subjected to experimental verification even if in principle. Given Kerr black 
holes are the only physically viable black holes, the study of Hawking effect
for the Kerr black holes in canonical formulation assumes additional 
importance.

In the section \ref{kerr-spacetime}, we begin with a brief discussion about the 
Kerr spacetime. In particular, we emphasize that unlike Schwarzschild spacetime 
a Kerr black hole spacetime has two horizons, of which the outer one is the 
event horizon. Then we discuss the properties of the corresponding null 
geodesics and null coordinates. 
In the section \ref{canonical-formulation}, we review the key aspects of the 
canonical formulation. We consider a minimally coupled massless scalar field in 
a Kerr black hole spacetime. Then we consider two asymptotic observers; one near 
past null infinity $\scriminus$ and another near future null infinity 
$\scriplus$. Following \cite{Barman:2017fzh}, we then define the pair of 
near-null coordinates to be used for canonical derivation. We then construct the 
Hamiltonian densities associated with the Fourier modes of the field as seen by 
these two observers.
In the section \ref{kerr-non-extremal}, we consider non-extremal Kerr 
black holes and then present the canonical derivation of the Hawking effect
as represented by the thermal distribution of the Hawking quanta. Subsequently, 
in the section \ref{kerr-extremal} we study the case for extremal Kerr black 
holes.

\section{The Kerr spacetime}\label{kerr-spacetime}

The spacetime geometry outside of a rotating black hole is described by the Kerr 
metric which is an exact vacuum solution of the Einstein equation in general 
relativity. It is further generalized by the advent of Kerr-Newman metric where 
one includes a net charge to a rotating black hole. However, it is rather a 
theoretical construct given a charged astrophysical body is unlikely to be found 
in nature. On the other hand the abundance of rotating Kerr black holes in our 
universe and the recent discovery of gravitational waves from their merger 
\cite{Abbott:2016blz, Abbott:2016nmj, TheLIGOScientific:2016pea, Abbott:2017vtc} 
makes them interesting astrophysical objects to investigate further.

\subsection{Metric and horizons in Kerr spacetime}

The Kerr spacetime is described by two parameters, namely the mass of the black 
hole $M$ and its angular momentum per unit mass $a$. Using the \emph{natural 
units} where speed of light $c$ and \emph{Planck constant} $\hbar$ are 
set to unity, one can express the corresponding invariant distance
element using \emph{Boyer-Lindquist} coordinates \cite{Boyer:1967} as
\begin{eqnarray}\label{eq:Kerr-Metric-in-Boyer-Lindquist}
ds^2 = -  \frac{1}{\rho^2} (\Delta -a^2\sin^2{\theta}) dt^2 + 
\frac{\rho^2}{\Delta} dr^2 + \rho^2 d\theta^2   \nonumber\\
~ + \frac{\Sigma}{\rho^2} \sin^2{\theta}~d\phi^2 - 
\frac{2a}{\rho^2} (r^2+a^2-\Delta) \sin^2{\theta} ~dt d\phi ~, 
\end{eqnarray}
where $\rho^2=r^2+a^2\cos^2{\theta}$, $\Sigma=(r^2+a^2)^2-a^2 
\Delta\sin^2{\theta}$ and $\Delta=r^2+a^2-r_{s}r$ with $r_{s}=2GM$ being the 
Schwarzschild radius corresponding to mass $M$ \cite{Kerr:2007dk, book:Poisson, 
Dadhich:2013qx, book:Schutz, book:raine, book:carroll2004, dInverno, 
book:PadmanabhanGrav, Heinicke:2015iva, Krasinski:1900zzb, Teukolsky:2014vca, 
Visser:2007fj, Smailagic:2010nv}. The metric components diverge at both 
$\rho^2=0$ and $\Delta=0$. In particular, the \emph{Kretschmann scalar} is 
singular  at $\rho^2=0$, which signifies a curvature singularity and cannot be 
removed by any coordinate transformation. On the other hand, $\Delta=0$ 
corresponds to a coordinate singularity and it gives the position of two 
horizons at $r=r_{h}$ and $r=r_{c}$ where
\begin{equation}\label{eq:horizons}
r_h= \tfrac{1}{2}(r_s + \sqrt{r_s^2 - 4 a^2} ~) ~,~
r_c= \tfrac{1}{2}(r_s - \sqrt{r_s^2 - 4 a^2} ~) ~.
\end{equation}
The outer horizon, located at $r_h$, is the event horizon with the surface 
gravity $\ksg_h=\sqrt{r_s^2-4a^2}/(2r_s r_h)$. The inner horizon is
located at $r_{c}$ and it is a Cauchy horizon with surface gravity  
$\ksg_c=\sqrt{r_s^2-4a^2}/(2r_s r_c)$.
Due to the frame-dragging effect \cite{book:Poisson,book:Schutz}, an inertial 
observer experiences an angular velocity in Kerr spacetime, given by
\begin{equation}\label{InertialAngularVelocity}
\Omega \equiv \Omega(r,\theta) = \frac{g^{t\phi}}{g^{tt}} 
= \frac{a r\rs}{\Sigma}   ~.
\end{equation}
We may mention that the study as presented here can be generalized for the 
Kerr-Newman black holes \cite{Newman:1965my, Adamo:2014baa, Newman:1965tw} by 
using $\Delta=r^2+a^2+r_Q^2-r_{s}r$, where $r_Q^2=Q^2G/4\pi\epsilon_0$ with 
charge $Q$ and Coulomb's force constant $1/4\pi\epsilon_0$.

\subsection{Null trajectories in Kerr spacetime}

In Kerr spacetime the governing equations for null geodesics 
\cite{book:Poisson} can be expressed as 
\begin{equation}\label{eq:null-geodesics}
 \dot{t} = \frac{r^2+a^2}{\Delta},~\dot{r} = \pm 1 ~,~ 
 \dot{\theta} = 0 ~,~ \dot{\phi} = \frac{a}{\Delta} ~,
\end{equation}
where the overhead dot denotes derivative with respect to an affine parameter.
Due to the frame dragging effect the azimuthal angle $\phi$ cannot be kept 
constant along any ingoing or outgoing null trajectory, unlike in Schwarzschild 
spacetime. However, using the Eqn. (\ref{eq:null-geodesics}) one can show that 
along the ingoing null trajectories, the coordinates $v=t+\rstar$ and 
$\psi=\phi+\rhash$ are constants where
\begin{equation}\label{eq:diff-tortoise}
 d\rstar=\frac{r^2+a^2}{\Delta}dr ~~,~~ d\rhash=\frac{a}{\Delta}dr~.
\end{equation}
Similarly, along the outgoing null trajectories the coordinates $u=t-\rstar$ 
and $\chi=\phi-\rhash$ are constants. Here $\rstar$ denotes the \emph{tortoise} 
coordinate in analogy to the one in Schwarzschild spacetime. Depending on 
whether the Kerr black hole is extremal or non-extremal, the expression of the 
coordinates $\rstar$ and $\rhash$ in terms of radial coordinate $r$ differ.

\subsection{Number density of Hawking quanta}

In order to study the Hawking effect we consider the scenario where Kerr 
spacetime is formed after the collapse of matters starting from a Minkowski 
spacetime in the past. The detailed evolution of the collapsing matters are not 
relevant for our study. To capture this aspect of change in metric over time, 
yet to avoid the technical difficulties that are associated with the field 
quantization in a single time-dependent metric, Hawking considered two different 
asymptotic observers, one at past null infinity $\scriminus$, and other at 
future null infinity $\scriplus$, each having time-independent but different 
metric. The vacuum states  corresponding to these two observers differ from each 
other as their metric are different.

To represent the Hawking quanta, in the given spacetime with metric 
$g_{\mu \nu}$, we consider a minimally coupled massless free scalar field 
$\Phi(x)$ which is described by the action 
\begin{equation}\label{eq:scalar-action-full}
S_{\Phi} = \int d^{4}x \left[ -\frac{1}{2} \sqrt{-g} g^{\mu \nu} 
\nabla_{\mu}\Phi(x) \nabla_{\nu}\Phi(x) \right]~.
\end{equation}
The Hawking effect is realized by computing the Bogoliubov transformation 
coefficients between these two observers at the past and the future  
null infinities respectively. The expectation value of the number density 
operator corresponding to the Hawking quanta of frequency $\omega$ is given by 
\cite{hawking1975}
\begin{equation}\label{NumberDensityOriginalHawking}
N_{\omega} = \frac{1}{e^{2\pi(\omega - m \Omega_h)/\ksg_{h}} - 1} ~,
\end{equation}
where $\ksg_{h}$ and $\Omega_h$ are the surface gravity and the angular 
velocity $\Omega$ at the event horizon respectively. Here $m$ denotes the 
azimuthal quantum number of the modes. By comparing the Eqn. 
(\ref{NumberDensityOriginalHawking}) with the blackbody distribution we may 
read off the corresponding Hawking temperature as $T_H = \ksg_{h}/(2\pi k_B)$ 
with $k_B$ being the Boltzmann constant.

\section{canonical formulation}\label{canonical-formulation}

The particle creation in a curved spacetime is directly connected to the 
dynamical nature of the spacetime metric. In the case of black hole radiation, 
it arises as the spacetime evolves from being Minkowskian in the past to a 
specific black hole spacetime in future due to the collapse of matters. In order 
to perform Hamiltonian based canonical derivation of the Hawking effect in Kerr 
spacetime we follow a similar approach by considering two asymptotic observers 
near past and future null infinities, each having time-independent but different 
metric. Subsequently, we compute expectation value of the Hamiltonian density 
operator for the field modes of the future observer in the vacuum state of the 
past observer and then read off the number density of the Hawking quanta.

\subsection{Reduced scalar field action}

In the Kerr spacetime with axial symmetry, one can decompose the scalar 
field in terms of spheroidal harmonics $e^{i m\phi}\mathscr{S}_{lm}(\theta)$ as 
$\Phi(x) = \sum_{l,m}e^{i m\phi}\mathscr{S}_{lm}(\theta)~ \varphi_{lm}(r,t) / 
\sqrt{r^2+a^2}$. However, in order to emphasize a key aspect of Kerr spacetime 
we perform the reduction in two steps. Firstly, we express the scalar field 
as $\Phi(t,r,\theta,\phi) = \sum_{lm} e^{i m\phi}~\Phi_{lm}(t,r,\theta)$. 
After carrying out the integration over azimuthal angle $\phi$, the action 
(\ref{eq:scalar-action-full}) reduces to $S_{\Phi} = \sum_{ll'm} S_{ll'm}$ where
\begin{eqnarray}
 S_{ll'm} &=& \int dt dr d\theta \sqrt{-g} \left[ 
-\tfrac{1}{2} g^{tt} \partial_t \Phi_{l'm}^* \partial_t \Phi_{lm}
 \right.\nonumber \\ && - ~~\left.
  \tfrac{i}{2} m~ g^{t\phi} 
(\partial_t \Phi_{l'm}^* \Phi_{lm} - \Phi_{l'm}^* \partial_t \Phi_{lm})
 \right.\nonumber \\ && \left. -~~\tfrac{1}{2} g^{rr} \partial_r 
\Phi_{l'm}^* \partial_r \Phi_{lm}-
\tfrac{1}{2} g^{\theta\theta} \partial_{\theta} \Phi_{l'm}^* \partial_{\theta} 
\Phi_{lm}\right.\nonumber\\
&& \left.~~~~~~~~~~~~~~~~~~~~~~~~ - 
\tfrac{1}{2} m^2 g^{\phi\phi} \Phi_{l'm}^*\Phi_{lm} \right] ~.
\label{eq:scalar-action-reduced-1}
\end{eqnarray}
We note that if one redefines the field further as
\begin{equation}\label{OmegaRedefinedScalarField}
\Phi_{lm}(t,r,\theta) \equiv e^{-i m \Omega t}~\Phit_{lm}(t,r,\theta)~,
\end{equation}
then the terms in the action (\ref{eq:scalar-action-reduced-1})
involving temporal derivative of fields simplify to
\begin{eqnarray}\label{eq:scalar-action-reduced-2}
 S_{ll'm} &=& \int dt dr d\theta \sqrt{-g} \left[ 
-\tfrac{1}{2} g^{tt} \partial_t \Phit_{l'm}^* \partial_t \Phit_{lm}
 \right.\nonumber \\ && - \left. 
\tfrac{1}{2} g^{rr} \partial_r (e^{-i m\Omega t}\Phit_{l'm})^* \partial_r 
(e^{-i m\Omega t}\Phit_{lm})
 \right.\nonumber \\ && - \left. 
\tfrac{1}{2} g^{\theta\theta} \partial_{\theta} (e^{-i m\Omega t}\Phit_{l'm})^* 
\partial_{\theta} (e^{-i m\Omega t}\Phit_{lm})\right.\nonumber\\
&& -  \left.
\tfrac{1}{2} m^2 \left(g^{\phi\phi}-\Omega g^{t\phi}\right) 
\Phit_{l'm}^* \Phit_{lm}  ~\right] ~.
\end{eqnarray}
In the regions near the past and the future null infinities where the relevant 
observers for realizing Hawking effect are located, the redefined field can be 
expressed as $\Phit_{lm}(t,r,\theta) \simeq \mathscr{S}_{lm}(\theta)~ 
\varphi_{lm} (\rstar,t)/\sqrt{r^2+a^2}$. The same approximation for the field is 
also possible in the region near the event horizon $\Delta\to 0$ where the term 
$\Omega$ becomes $\Omega_{h}$ which is the angular velocity of the event 
horizon.
By using orthogonality condition $\int d(\cos{\theta}) \mathscr{S}_{lm} 
(\theta) 
\mathscr{S}^*_{l'm}(\theta) = \delta_{l,l'}$, we achieve the final form of 
the reduced action as $S_{\Phi} = \sum_{lm} S_{lm}$ in the regions near horizon 
as well as near null infinities, where 
\begin{equation}\label{eq:2d-reduced-scalar-action}
 S_{lm} \simeq \int dt d\rstar 
\left[\tfrac{1}{2}
\partial_t \varphi^*_{lm}\partial_t\varphi_{lm} 
- \tfrac{1}{2}
\partial_{\rstar}\varphi^*_{lm}\partial_{\rstar}\varphi_{lm}
 \right] ~.
\end{equation}
The action (\ref{eq:2d-reduced-scalar-action}) represents a scalar field in 
1+1 dimensional flat spacetime.

\subsection{Frequency shift due to frame dragging}

The solutions to the field equation corresponding to the action 
(\ref{eq:2d-reduced-scalar-action}) can be expressed as
\begin{equation}\label{eq:scalar-field-asymptotic-solutions}
\varphi_{lm}(r,t) \sim \frac{1}{\sqrt{2\pi \omegat }} 
e^{-i\omegat (t\pm r_{\star})}~.
\end{equation}
However, in order to understand the full dynamics of the physical field 
$\Phi$, one needs to consider the solutions 
(\ref{eq:scalar-field-asymptotic-solutions}) together with the relation 
(\ref{OmegaRedefinedScalarField}) which provides additional time-dependence.
In particular, if one reads off the frequency, as defined as the eigenvalue of 
the operator $i\partial_t$, then it would be $\omegat$ for redefined 
field mode $\varphi_{lm}$ (\ref{eq:scalar-field-asymptotic-solutions}). On the 
other hand, the frequency, say $\omega$, of the physical field mode $\Phi_{lm}$ 
(\ref{OmegaRedefinedScalarField}) would be $\omega = \omegat + m \Omega$.
The Hawking effect is realized through the modes which travel out from the 
region very close to the event horizon. Therefore, for these modes the 
frequency $\omegat$ can be related to the physical frequency $\omega$ as 
\cite{Ford:1975tp, Hod:2011zzd, Menezes:2016SeaKs, Menezes:2017oeb, Lin:2009wm, 
Miao:2011dy, Iso:2006ut}
\begin{equation}\label{eq:FrequencyShift}
\omegat = \omega - m\Omega_h ~. 
\end{equation}
This key feature of frequency shift in the Kerr spacetime is reflected 
through the expression of the expectation value of the number density operator 
(\ref{NumberDensityOriginalHawking}).

\subsection{The observers $\observerminus$ and $\observerplus$}

\subsubsection{Near-null coordinates}

\begin{figure} 
\includegraphics[width=0.45\textwidth]{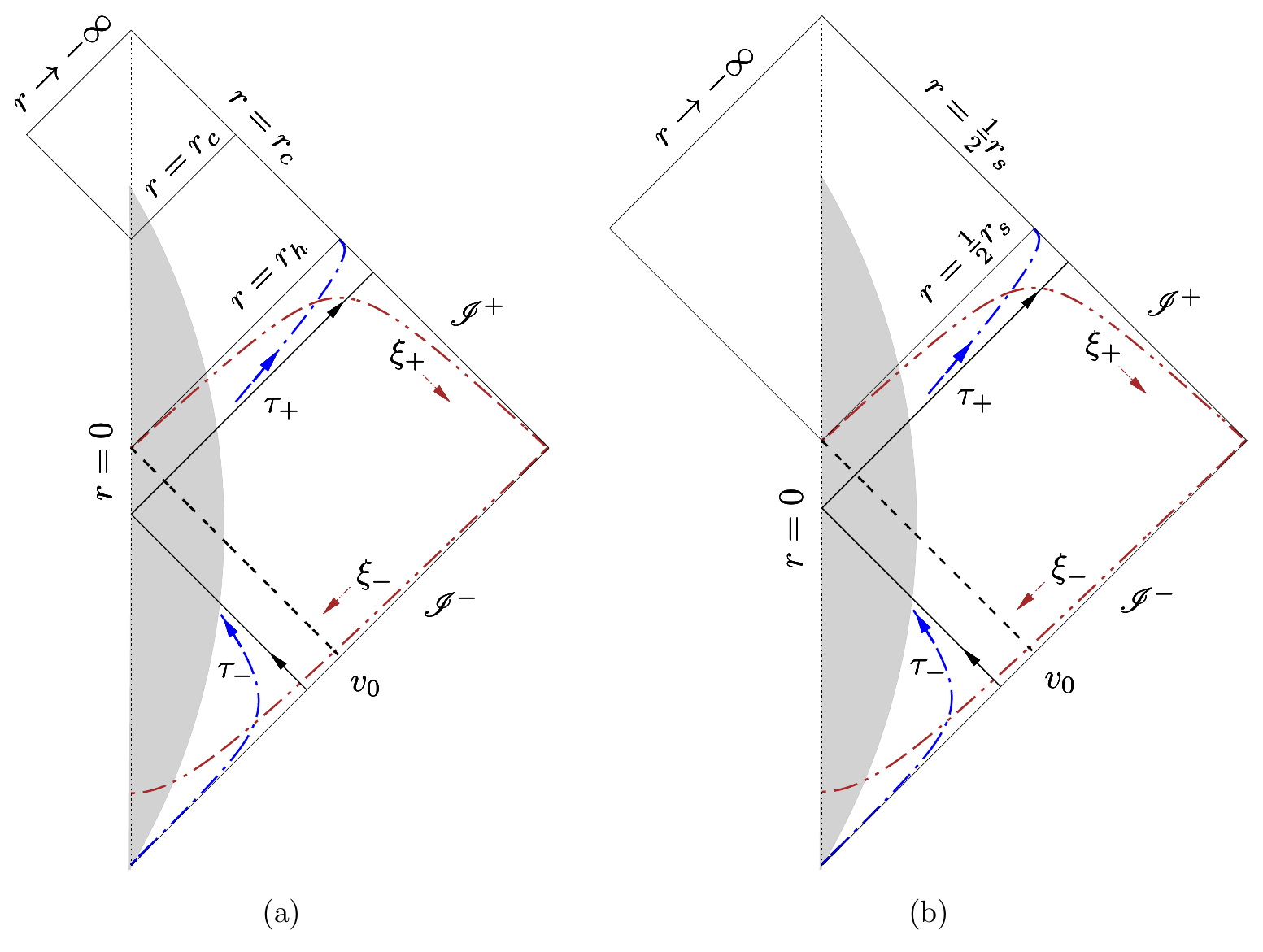}
\caption{Simplified Penrose diagrams for (a) non-extremal and (b) extremal Kerr 
black holes. The black hole is formed through matter collapse which is 
depicted by the shaded region. A null ray leaving from $\scriminus$ 
before $v_0$ would end up on $\scriplus$ whereas a null ray leaving after 
$v_{0}$ would end up being inside the black hole.}
\label{fig:KerrPenroseDiagram}
\end{figure}

The field modes (\ref{eq:scalar-field-asymptotic-solutions}) are usually 
expressed in terms of the null coordinates $v$ and $u$ as 
$\tilde{\varphi}_{lm} \sim e^{-i\omegat v}$ or $\tilde{\varphi}_{lm} \sim 
e^{-i\omegat u}$. Therefore, the Hawking effect is conveniently understood 
using Bogoliubov transformation coefficients between field modes of the two 
observers, each are described by null coordinates (see 
Fig.\ref{fig:KerrPenroseDiagram}). However, the usage of the 
null coordinates do not lead to a true matter Hamiltonian that can describe the 
dynamics of these modes. Therefore, in the pursuit of a canonical derivation of 
the Hawking effect we need to look for coordinates which are not null. By 
following the approach as prescribed in \cite{Barman:2017fzh} we define 
a set of \emph{near-null} coordinates by slightly deforming the outgoing and the
ingoing null coordinates. In particular, for the observer located near the past 
null infinity $\scriminus$, say observer $\observerminus$, the near-null 
coordinates are defined as
\begin{equation}\label{NearNullCoordinatesMinus}
\tau_{-} = t - (1-\epsilon)\rstar ~~;~~ \xi_{-} = -t - (1+\epsilon)\rstar  ~,
\end{equation}
where the parameter $\epsilon$ is considered to be small such that 
$\epsilon\gg\epsilon^2$. In a similar manner we define the near-null 
coordinates for the observer located near the future null infinity $\scriplus$, 
say observer $\observerplus$, as
\begin{equation}\label{NearNullCoordinatesPlus}
\tau_{+} = t + (1-\epsilon)\rstar ~~;~~ \xi_{+} = -t + (1+\epsilon)\rstar  ~.
\end{equation}
We are considering the scenario where the black hole is formed after the 
collapse of matters staring from a Minkowskian spacetime. Therefore, for the 
past observer $\observerminus$, the definition of tortoise coordinate in 
(\ref{NearNullCoordinatesMinus}) is trivial \emph{i.e.} $d\rstar = dr$. We may 
note that the timelike characteristics of the coordinates $\tau_{\pm}$ is 
maintained for the range $0<\epsilon<2$. However, for simplicity here we 
consider the parameter $\epsilon$ to be small.

\subsubsection{Field Hamiltonian}

For the past observer $\observerminus$, the 1+1 dimensional reduced spacetime is 
described by the Minkowski metric $ ds^2 = -dt^2 + d\rstar^2 = -dt^2 + dr^2$. 
Therefore, the invariant line element can be written using near-null 
coordinates (\ref{NearNullCoordinatesMinus}) as
\begin{equation}\label{NearNullMetricMinus}
ds^2_{-} = \tfrac{\epsilon}{2} [ - d\tau_{-}^2 + 
d\xi_{-}^2 +\tfrac{2}{\epsilon} d\tau_{-} d\xi_{-} ] 
\equiv \tfrac{\epsilon}{2}~ g^{0}_{\mu\nu}dx_{-}^{\mu} dx_{-}^{\nu}  ~ ,
\end{equation}
where flat metric $g^{0}_{\mu\nu}$ is conformally transformed. With respect to 
the future observer $\observerplus$, the Kerr black hole is already formed. 
Nevertheless, as far as the dynamics of the 1+1 dimensional reduced scalar field 
action (\ref{eq:2d-reduced-scalar-action}) is concerned, even for the observer 
$\observerplus$, the underlying metric can be expressed as $ ds^2 =  -dt^2 + 
d\rstar^2$. Using the near-null coordinates (\ref{NearNullCoordinatesPlus}), 
this invariant line-element becomes
\begin{equation}\label{NearNullMetricPlus}
ds^2_{+} = \tfrac{\epsilon}{2} [ - d\tau_{+}^2 + 
d\xi_{+}^2 +\tfrac{2}{\epsilon} d\tau_{+} d\xi_{+} ]
\equiv  \tfrac{\epsilon}{2}~g^{0}_{\mu\nu}dx^{\mu}_{+}dx^{\nu}_{+} ~.
\end{equation}
Therefore, we may express the reduced scalar field action 
(\ref{eq:2d-reduced-scalar-action}) for both observers as 
\begin{equation}\label{ReducedScalarAction2DFlatMinus}
S_{\varphi} =  \int d\tau_{\pm}  d\xi_{\pm} \left[-\tfrac{1}{2} \sqrt{-g^{0}} 
g^{0\mu\nu} \partial_{\mu}\varphi \partial_{\nu} \varphi \right]  ~.
\end{equation}
For brevity of notation we have omitted the subscripts from the redefined field 
$\varphi_{lm}$. 

In order to derive the scalar field Hamiltonian, we consider spatial slicing of 
the reduced spacetime labelled by the coordinate $\tau_{\pm}$. From Eqns. 
(\ref{NearNullMetricMinus}) and (\ref{NearNullMetricPlus}), one can show that 
corresponding \emph{lapse function} $N = 1/\epsilon$, \emph{shift vector} $N^1 = 
1/\epsilon$ and determinant of the spatial metric $q = 1$. The scalar field 
Hamiltonian then can be written as
\begin{equation}\label{ScalarHamiltonianOpm}
H_{\varphi}^{\pm} = \int d\xi_{\pm}\; \tfrac{1}{\epsilon}\left[\left\{ 
\tfrac{1}{2} \Pi^2  + \tfrac{1}{2}  (\partial_{\xi_{\pm}}\varphi)^2 \right\} 
+ \Pi~ \partial_{\xi_{\pm}} \varphi \right] ~,
\end{equation}
where the superscript $(\pm)$ refers to the Hamiltonian for the observer 
$\observerplus$ and $\observerminus$ respectively. The field $\varphi$ and its 
conjugate momentum $\Pi$ satisfy the Poisson bracket 
\begin{equation}\label{PoissonBracketPM}
 \{\varphi(\tau_{\pm},\xi_{\pm}), \Pi(\tau_{\pm},\xi_{\pm}')\} = 
\delta(\xi_{\pm} - \xi_{\pm}') ~.
\end{equation}
Using Hamilton's equation, the field momentum $\Pi$ can be expressed as
\begin{equation}\label{FieldMomentumPM}
\Pi(\tau_{\pm},\xi_{\pm}) = \epsilon\;\partial_{\tau_{\pm}}\varphi - 
\partial_{\xi_{\pm}}\varphi ~.
\end{equation}
We note from the Eqn. (\ref{ScalarHamiltonianOpm}) that at the value of the 
parameter $\epsilon=0$, the Hamiltonian becomes ill-defined. This 
signifies the necessity of near-null coordinates in order to study the Hawking 
effect using a Hamiltonian approach.

\subsubsection{Fourier modes}

For both the observers, the spatial volume $V_{\pm}=\int d\xi_{\pm}\sqrt{q}$ 
formally diverges as $\sqrt{q}=1$. Therefore to avoid dealing with explicitly 
diverging quantities, we consider a finite fiducial box during the intermediate 
steps of computations, such that
\begin{equation}\label{SpatialVoumePM}
V_{\pm} = \int_{\xi_{\pm}^L}^{\xi_{\pm}^R} d\xi_{\pm}\sqrt{q} = {\xi_{\pm}^R} - 
{\xi_{\pm}^L}  ~.
\end{equation}
Subsequently, we define the respective Fourier modes of the scalar field 
for the observers $\observerplus$ and $\observerminus$ as
\begin{eqnarray}
\varphi(\tau_{\pm},\xi_{\pm}) &=& \tfrac{1}{\sqrt{V_{\pm}}}\sum_{k} 
\tilde{\phi}^{\pm}_{k} 
e^{i k \xi_{\pm}}  ~,
\nonumber \\
\Pi(\tau_{\pm},\xi_{\pm}) &=&  \tfrac{1}{\sqrt{V_{\pm}}} \sum_{k} \sqrt{q}~ 
\tilde{\pi}^{\pm}_{k} 
e^{i k \xi_{\pm}} ~,
\label{FourierModesDefinitionPM}
\end{eqnarray}
where $\tilde{\phi}^{\pm}_{k} = \tilde{\phi}^{\pm}_{k} (\tau_{\pm})$, 
$\tilde{\pi}^{\pm}_{k} = \tilde{\pi}^{\pm}_{k} (\tau_{\pm})$ are the 
complex-valued mode functions. The finite volume of the fiducial box leads to 
the definition of Kronecker delta and Dirac delta as $\int d\xi_{\pm}\sqrt{q}  
e^{i(k-k')\xi_{\pm}} = V_{\pm} \delta_{k,k'}$ and $\sum_k 
e^{ik(\xi_{\pm}-\xi_{\pm}')} = V_{\pm} \delta(\xi_{\pm}-\xi_{\pm}')/\sqrt{q}$ .
The definition of these two deltas together imply $k\in \{k_s\}$ where 
$k_s=2\pi s/V_{\pm}$ with $s$ being a nonzero integer. These definitions help us 
to express the scalar field Hamiltonians (\ref{ScalarHamiltonianOpm}) in terms 
of the Fourier modes as $H_{\varphi}^{\pm} = \sum_{k} \tfrac{1}{\epsilon}
\left(\mathcal{H}_{k}^{\pm} + \mathcal{D}_{k}^{\pm}\right)$ where the 
Hamiltonian densities and diffeomorphism generators are
\begin{equation}\label{FourierHamiltonianPM}
\mathcal{H}_{k}^{\pm} = \tfrac{1}{2} \tilde{\pi}^{\pm}_{k}  
\tilde{\pi}^{\pm}_{-k} + \tfrac{1}{2} k^2 \tilde{\phi}^{\pm}_{k} 
\tilde{\phi}^{\pm}_{-k} ~,
\end{equation}
and
\begin{equation}\label{DiffeomorphismGenerator}
\mathcal{D}_{k}^{\pm} =  
 -\tfrac{i k}{2} ( \tilde{\pi}^{\pm}_{k} \tilde{\phi}^{\pm}_{-k} -
 \tilde{\pi}^{\pm}_{-k} \tilde{\phi}^{\pm}_{k} )  ~,
\end{equation}
respectively. The corresponding Poisson brackets are
\begin{equation}\label{FourierPoissonBracketPlus}
\{\tilde{\phi}^{\pm}_{k}, \tilde{\pi}^{\pm}_{-k'}\} = \delta_{k,k'} ~.
\end{equation}

\subsection{Relation between Fourier modes}
       
In order to find relations between the field modes and their conjugate momenta 
for the two observers, we note that $\varphi(\tau_{-},\xi_{-}) = 
\varphi(\tau_{+},\xi_{+})$, given the field is scalar. The field momentum 
follows a relation $\Pi(\tau_{+},\xi_{+}) = (\partial \xi_{-}/\partial \xi_{+}) 
\Pi(\tau_{-},\xi_{-})$ \cite{Barman:2017fzh}. A simple way to understand this 
relation is as follows. In order to realize the Hawking effect, the past 
observer considers ingoing modes with null coordinate $v$ constant 
whereas the future observer considers the outgoing modes with null 
coordinate $u$ constant. Using these restrictions together with expressions 
of the momenta (\ref{FieldMomentumPM}), one can arrive at the given 
relation between the two momenta. Having these relations between the field and 
the field momentum, one can obtain the relations between their
Fourier modes and respective conjugate momenta as
\begin{equation}\label{FieldModesRelation}
\tilde{\phi}^{+}_{\kr} = \sum_{k} \tilde{\phi}^{-}_{k} F_{0}(k,-\kr) ~;~
\tilde{\pi}^{+}_{\kr} =  \sum_{k} \tilde{\pi}^{-}_{k} F_{1}(k,-\kr)  ~,
\end{equation}
where we have considered Fourier modes on fixed spatial hyper-surfaces. The 
coefficient functions $F_{n}(k,\kr)$ are given by
\begin{equation}\label{FFunctionGeneral}
F_{n}(k,\kr) = \frac{1}{\sqrt{V_{-} V_{+}}} \int d\xi_{+} 
\left(\tfrac{\partial \xi_{-}}{\partial \xi_{+}}\right)^n
~e^{i k \xi_{-} + i \kr \xi_{+}} ~,
\end{equation}
where $n=0,1$. These coefficient functions are analogous to the Bogoliubov 
coefficients in the covariant formulation. In particular we note that for $k,\kr 
>0$ the coefficient functions $F_{n}(-k,-\kr)$ are analogous to the Bogoliubov 
mixing coefficients $\beta_{\omega\omega'}$ whereas $F_{n}(k,-\kr)$ are 
analogous to the Bogoliubov coefficients $\alpha_{\omega\omega'}$ of 
\cite{hawking1975}. Using representation of Dirac delta distribution 
$\delta(\mu) = \tfrac{1}{2\pi} \int dx~e^{i\mu x}$ and by setting $\mu=1$, 
$x=(\pm k\xi_{-}+\kr\xi_{+})$ there one can obtain a relation 
\begin{equation}\label{F0F1Relation}
F_{1}(\pm k,\kr) = \mp \left(\tfrac{\kr}{k}\right) F_{0}(\pm k,\kr) ~.
\end{equation}
In other words, the evaluation of only one coefficient function, say 
$F_{0}(\pm k,\kr)$, is sufficient for the subsequent analysis.

\subsection{Poisson bracket consistency condition}

The requirement that two different Poisson brackets 
$\{\tilde{\phi}^{-}_{k},\tilde{\pi}^{-}_{-k'}\} = \delta_{k,k'}$ and 
$\{\tilde{\phi}^{+}_{\kr},\tilde{\pi}^{+}_{-\kr'}\} = \delta_{\kr,\kr'}$ be 
simultaneously satisfied, demands a relation between the coefficient 
functions $F_0(\pm k,\kr)$. In particular, by using the Eqn. 
(\ref{F0F1Relation}), we may express this consistency requirement as
\begin{equation}\label{PoissonBracketConsistencyCondition}
\mathbb{S}_{-}(\kr) - \mathbb{S}_{+}(\kr) = 1 ~,
\end{equation}
where $\mathbb{S}_{\pm}(\kr) = \sum_{k>0} (\kr/k) |F_{0}(\pm k,\kr)|^2$. This 
condition is analogous to the consistency condition between Bogoliubov 
coefficients \cite{Alvarenga:2003tx} which arises from the imposition of the 
commutator bracket between the creation and annihilation operators of the field 
modes for two asymptotic observers.

\subsection{Relation between Hamiltonian densities and diffeomorphism 
generators}

Using relations (\ref{FieldModesRelation}) and (\ref{F0F1Relation}) one can 
express the Hamiltonian density $\mathcal{H}_{\kr}^{+}$ for the observer 
$\observerplus$ in terms of the Hamiltonian density $\mathcal{H}_{k}^{-}$ of the 
observer $\observerminus$ as 
\begin{equation}\label{ModesHamiltonianRelations0}
\mathcal{H}_{\kr}^{+} = h_{\kr}^1 + \sum_{k>0} \left(\frac{\kr}{k}\right)^2 
\left[\right|F_{0}(-k,\kr)|^2 + |F_{0}(k,\kr)|^2] ~\mathcal{H}_k^{-}  ~,
\end{equation}
where 
$h_{\kr}^1 =  \sum_{k\neq k'} (\kr^2/2 k k') F_{0}(k,-\kr) F_{0}(-k',\kr) 
\{\tilde{\pi}^{-}_{k} \tilde{\pi}^{-}_{-k'} + kk' \tilde{\phi}^{-}_{k} 
\tilde{\phi}^{-}_{-k'} \}$. $h_{\kr}^1$ is being linear in $\phi^{-}_{k}$ and 
its conjugate momentum, the vacuum expectation value of its quantum counterpart 
vanishes. Similarly, the diffeomorphism generators of the two observers can be 
related as
\begin{equation}\label{ModesDiffeomorphismRelations0}
\mathcal{D}_{\kr}^{+} = d_{\kr}^1 + \sum_{k>0} \left(\frac{\kr}{k}\right)^2 
\left[\right|F_{0}(-k,\kr)|^2 + |F_{0}(k,\kr)|^2] ~\mathcal{D}_k^{-}  ~,
\end{equation}
where  $ d_{\kr}^1 = \sum_{k\neq k'} (i\kr^2/2k)~ \{F_{0}(-k,\kr) 
F_{0}(k',-\kr)~ \tilde{\pi}^{-}_{-k}  \tilde{\phi}^{-}_{k'} - F_{0}(k,-\kr) 
F_{0}(-k',\kr)~ \tilde{\pi}^{-}_{k} \tilde{\phi}^{-}_{-k'} \}$ which is also 
linear in field mode and its conjugate momentum.

\subsection{Fock quantization and the vacuum state}

The scalar field under consideration is real-valued which imposes condition on 
the Fourier modes as $\tilde{\phi}^{*}_{k}=\tilde{\phi}_{-k}$. This implies that 
the real and imaginary parts of field modes are not independent. A suggested way 
to implement this reality condition is to suitably redefine real and imaginary 
parts for different domains in terms of a real-valued mode function 
\cite{Hossain:2010eb,Barman:2017fzh} which leads the Hamiltonian density to 
represent a simple harmonic oscillator as
\begin{equation}\label{RealHamiltonian}
\mathcal{H}_{k}^{\pm} = \frac{1}{2} \pi^{2}_{k}
+ \frac{1}{2} k^2 \phi^{2}_{k} 
~,~~~\{\phi^{2}_{k},\pi^{2}_{k'}\}=\delta_{k,k'}~,
\end{equation}
where $\phi_{k}$ and $\pi_{k}$ are the redefined real-valued field modes. 
Further, this redefinition makes diffeomorphism generator to vanish \emph{i.e.} 
$\mathcal{D}_k^{-} = 0$.

The Fock quantization of \emph{massless free} scalar field can be viewed as the  
Schr\"odinger quantization of only \emph{positive frequency} oscillator modes.
We may now restrict ourselves with the modes where $k,\kr > 0$ so that the 
mode frequency can be identified as $\omegat = \kr$ and so on. The energy 
spectrum for each of these oscillator modes is given by 
$\hat{\mathcal{H}}_{k}^{-}|n_k\rangle = (\hat{N}_{k}^{-}+\tfrac{1}{2}) k 
|n_k\rangle = (n+\tfrac{1}{2}) k |n_k\rangle $ where $\hat{N}_{k}^{-}$ is the 
corresponding number operator, $|n_k\rangle$ are its eigen-states with integer 
eigenvalues $n \ge 0$. The Hawking effect is realized by computing the 
expectation value of the Hamiltonian density operator 
$\hat{\mathcal{H}}_{\kr}^{+} \equiv (\hat{N}_{\kr}^{+}+\tfrac{1}{2})$ 
corresponding to the observer $\observerplus$ in the vacuum state $|0_{-}\rangle 
= \Pi_{k} |0_k\rangle$ corresponding to the observer $\observerminus$. 
Therefore, the expectation value of the number density operator corresponding to 
the Hawking quanta of frequency $\omegat = \kr$, after using the Eqn. 
(\ref{PoissonBracketConsistencyCondition}) along with the Eqn. 
(\ref{ModesHamiltonianRelations0}), can be expressed as
\begin{equation}\label{NumberDensityEVGeneral} 
N_{\omegat} = N_{\kr} \equiv \langle 0_{-}|\hat{N}_{\kr}^{+}|0_{-}\rangle 
=  \mathbb{S}_{+}(\kr)   ~,
\end{equation}
where we have used the properties $\langle 0_k|\hat{\phi}_{k} 
|0_k\rangle = 0$ and $\langle 0_k|\hat{\pi}_{k} |0_k\rangle = 0$. For Fock 
quantization, the number density operator  employed in \cite{Barman:2017fzh}  
is equivalent to the number density operator (\ref{NumberDensityEVGeneral}).

\section{Non-extremal Kerr black holes}\label{kerr-non-extremal}

In order to explicitly evaluate the coefficient function $F_0(k,\kr)$ we require 
the expression of tortoise coordinate $\rstar$ which depends crucially on the 
fact whether the given Kerr black hole is extremal or non-extremal. Therefore, 
we deal with these two cases separately. Using the Eqn. (\ref{eq:diff-tortoise}) 
one can compute the expression of $\rstar$ for non-extremal black hole,
with suitable choice of integration constants, as
\begin{equation}\label{eq:tortoise-non-extremal}
 \rstar=r+\frac{1}{2\ksg_{h}} \ln \left[(r-r_{h})\ksg_{h}\right] -
\frac{1}{2\ksg_{c}} \ln \left[(r-r_{c})\ksg_{c}\right]~,
\end{equation}
where $\ksg_{h}$ and $\ksg_{c}$ denote the surface gravity at the outer and 
the inner horizon of the Kerr spacetime respectively.

\subsection{Relation between spatial coordinates $\xi_{-}$ and $\xi_{+}$}

In order to establish the relation between the coordinates $\xi_{-}$ 
and $\xi_{+}$, following \cite{Barman:2017fzh}, we consider a pivotal 
point $\xi_{-}^0$ on a $\tau_{-} = constant$ hyper-surface. A spacelike 
interval on this hyper-surface can be written as
\begin{equation}\label{ScriMinusXiMinuxInterval}
(\xi_{-} - \xi_{-}^0)_{|\tau_{-}} = 2(\rstar^0-\rstar)_{|\tau_{-}} =
2(r^0-r)_{|\tau_{-}} \equiv \Delta ~, 
\end{equation}
where $r^0$ is a pivotal value corresponding to $\xi_{-}^0$. In deriving Eqn. 
(\ref{ScriMinusXiMinuxInterval}) we have used fact that for the observer 
$\observerminus$ the spacetime was Minkowskian. In a similar manner we can 
express a spacelike interval on a $\tau_{+} = constant$ hyper-surface as
\begin{equation}\label{ScriMinusXiPlusInterval}
(\xi_{+} - \xi_{+}^0)_{|\tau_{+}} 
= \Delta + \frac{1}{\ksg_{h}} \ln \left(1 + \frac{\Delta}{\Delta_{h}}\right) -
\frac{1}{\ksg_{c}} \ln \left(1 + \frac{\Delta}{\Delta_{c}}\right)~,
\end{equation}
where $\Delta_{h} \equiv 2(r^0 - r_{h})_{|\tau_{+}}$, $\Delta_{c} \equiv 2(r^0 
- r_{c})_{|\tau_{+}}$. Further, we have identified the interval $2(r 
- r^0)_{|\tau_{+}}$ as $\Delta$ using \emph{geometric optics approximation}. We 
choose the pivotal values $\xi_{-}^0 = \Delta_{h}$ and $\xi_{+}^0 = \xi_{-}^0 + 
\frac{1}{\ksg_{h}} \ln (\ksg_{h} \xi_{-}^0)-\frac{1}{\ksg_{c}} \ln 
(1+\ksg_{h}\xi_{-}^0/\sigma)$. These choices lead to the relation 
\begin{equation}\label{eq:near-null-ralation}
 \xi_{+} = \xi_{-} + \frac{1}{\ksg_{h}} \ln (\ksg_{h}\xi_{-}) - 
\frac{1}{\ksg_{c}} \ln \left(1 + \frac{\ksg_{h}\xi_{-}}{\sigma}\right) ~,
\end{equation}
where $\sigma=\ksg_{h}(\Delta_{c} -\Delta_{h})$. The modes that give rise to 
the Hawking radiation, travel out from the region very close to the horizon and 
for them $\ksg_{h}\xi_{-} \ll 1$. Consequently for these modes, the relation 
(\ref{eq:near-null-ralation}) can be approximated as
\begin{equation}\label{eq:final-near-null-relation}
  \xi_{+} \approx \tfrac{1}{\ksg_{h}} \ln (\ksg_{h}\xi_{-})  ~.
\end{equation}
We note from the Eqn. (\ref{eq:final-near-null-relation}) that the full domain 
of the coordinate $\xi_{+}$ is $(-\infty,\infty)$ whereas it is $(0,\infty)$ 
for $\xi_{-}$ \emph{i.e.} the domains are the same as implied by the Eqn. 
(\ref{eq:near-null-ralation}). However, as mentioned earlier, we shall restrict
ourselves within a finite fiducial box during the intermediate steps in our 
analysis.

\subsection{Evaluation of coefficient functions $F_{0}(\pm k,\kr)$}

From the Eqns. (\ref{PoissonBracketConsistencyCondition}) and 
(\ref{ModesHamiltonianRelations0}) we observe that the consistency 
condition and the Hamiltonian density both require the expression  of 
$F_{0}(k,\kr)$ and for non-extremal Kerr black hole it can be written as 
\begin{equation}\label{F0NonExtremal}
F_{0}(\pm k,\kr) = \int \frac{d\xi_{-}}{\sqrt{V_{-} V_{+}}}
(\ksg_{h}\xi_{-})^{-1}
e^{\pm i k \xi_{-} + i (\kr/\ksg_{h})\ln (\ksg_{h}\xi_{-})}  ~.
\end{equation}
The integrand being oscillatory in nature, the coefficient function 
$F_{0}(k,\kr)$ (\ref{F0NonExtremal}) is formally divergent. In order to 
regulate this integral we introduce the standard `$i\delta$' regulator, with 
small $\delta >0$, as follows
\begin{eqnarray}\label{F0NonExtremalRegulated}
 F_{0}^{\delta}(\pm k,\kr) &=& 
\int \frac{d\xi_{-}}{\sqrt{V_{-} V_{+}}}
(\ksg_{h}\xi_{-})^{-1} 
~ e^{-(\delta \mp i) k \xi_{-}}  \nonumber \\ 
&& ~ \times ~ e^{(\delta + i\kr/\ksg_{h} )\ln (\ksg_{h}\xi_{-})} ~.
\end{eqnarray}
In the limit $\delta\to 0$, the regulated expression $F_{0}^{\delta}(\pm 
k,\kr)$  reduces to $F_{0}(\pm k,\kr)$. We may mention that the regularization 
scheme employed in \cite{Barman:2017fzh} differs slightly from the one used 
here. By introducing variables $b_{\pm} = (\delta \mp i) k/\ksg_{h}$, 
$b_0 = (\delta + i\kr/\ksg_{h})$ and $\xi = (b_{\pm} \ksg_{h}\xi_{-})$, we can 
express regulated coefficient function as
\begin{equation}\label{F0NonExtremalEvaluated}
 F_{0}^{\delta}(\pm k,\kr) = 
 \frac{b_{\pm}^{-b_0}}{\ksg_{h} \sqrt{V_{-} V_{+}}} 
\int d\xi ~e^{-\xi} ~\xi^{b_0-1}  
= \frac{b_{\pm}^{-b_0} \Gamma(b_0)}{\ksg_{h} \sqrt{V_{-} V_{+}}} ~,
\end{equation}
where $\Gamma(b_0)$ is the Gamma function. Given the fiducial box has a finite 
volume, we have added two boundary terms $\Delta I^{L} = \int_0^{\xi^L} d\xi 
e^{-\xi} \xi^{b_0-1}$ and $\Delta I^{R} = \int_{\xi^R}^{\infty} d\xi e^{-\xi} 
\xi^{b_0-1}$ to make the Gamma function complete. Both of these terms vanish 
when one removes the volume regulators by taking the limit $\xi^L \equiv 
(b_{\pm}\ksg_{h}\xi_{-}^L)\to 0$ and $\xi^R \equiv 
(b_{\pm}\ksg_{h}\xi_{-}^R)\to\infty$. We note an useful property
\begin{equation}\label{F0F0Relation}
F_{0}^{\delta}(-k,\kr) = e^{(\pi-2\delta)\kr/\ksg_{h} - 
i\delta\pi}~F_{0}^{\delta}(k,\kr)  ~,
\end{equation}
where we have used $(\delta \pm i) = e^{\pm i(\pi/2 -\delta)} + 
\mathcal{O}(\delta^2)$. The Eqn. (\ref{F0F0Relation}) shows that
these coefficient functions satisfy a relation analogous to the Bogoliubov 
coefficients \cite{hawking1975} for non-extremal Kerr black hole.

\subsection{Consistency condition}          

The Eqn. (\ref{F0NonExtremalEvaluated}) together with the relation 
$k := k_s = (2\pi s/V_{-})$ leads
\begin{equation}\label{SplusNonExtremal1}
\mathbb{S}^{\delta}_{+}(\kr) = \frac{\kr ~|\Gamma(b_0)|^2 
e^{-(\pi-2\delta)\kr/\ksg_{h}}}
{\ksg_{h}^{2-2\delta} (2\pi)^{1+2\delta}} ~ 
\left(\frac{\zeta(1+2\delta)}{V_{-}^{-2\delta} V_{+}} \right) ~,
\end{equation}
where $\zeta(1+2\delta) = \sum_{s=1}^{\infty} s^{-(1+2\delta)}$ is the 
\emph{Riemann zeta function}. Furthermore, the Eqn. (\ref{F0F0Relation}) 
implies that $\mathbb{S}^{\delta}_{-}(\kr) = e^{(2\pi-4\delta)\kr/\ksg_{h}} 
~\mathbb{S}^{\delta}_{+}(\kr)$. Given $\zeta(1)$ is divergent, it is clear that  
in order to keep the term $\mathbb{S}^{\delta}_{\pm}$ finite one needs to 
remove volume regulators $\xi_{-}^L$ and $\xi_{-}^R$ along with the integral 
regulator $\delta$. To find the required dependency among the regulators, we use 
the regulated expression (\ref{F0NonExtremalEvaluated}) such that the 
consistency condition (\ref{PoissonBracketConsistencyCondition}) becomes
\begin{equation}\label{PoissonBracketConsistencyCondition2}
\frac{\sinh((\pi-2\delta)\kr/\ksg_{h})}
{\pi~(\kr/\ksg_{h})^{-1} |\Gamma(b_0)|^{-2}} = 
\frac{(\ksg_{h}V_{+}) (2\pi/\ksg_{h}V_{-})^{2\delta} }{\zeta(1+2\delta)} ~.
\end{equation}
Using Gamma function identity $\Gamma(z)\Gamma(1-z) = \pi/\sin\pi z$, zeta 
function identity $\lim_{\delta\to0}[\delta~ \zeta(1+\delta)] = 1$ and the 
Eqn. (\ref{eq:final-near-null-relation}) one can show that the consistency 
condition demands $\ksg_{h} \xi_{-}^L \sim e^{-1/2\delta}$, i.e. the volume 
regulator $\xi_{-}^L$ and integral regulator $\delta$ should be varied together. 
Once this limit is taken other volume regulator $\xi_{-}^R$ drops off from the 
expression of $\mathbb{S}^{\delta}_{+}(\kr)$.

\subsection{Number density of Hawking quanta}

Therefore, the expectation value of the number density operator 
(\ref{NumberDensityEVGeneral}) for a non-extremal Kerr black hole becomes
\begin{equation}\label{NumberDensityEVNonExtremal} 
N_{\kr} = \lim_{\delta \to 0} \mathbb{S}^{\delta}_{+}(\kr) 
= \frac{1}{e^{2\pi\kr/\ksg_{h}} - 1} ~. 
\end{equation}
The wave number $\kr$ in the Eqn. (\ref{NumberDensityEVNonExtremal}) 
corresponds to the redefined field $\varphi_{lm}$. Therefore, following the 
relation (\ref{eq:FrequencyShift}) together with $\tilde{\omega}=\kr >0$, the 
number density of Hawking quanta of frequency $\omega$
corresponding to the physical field mode $\Phi_{lm}$ becomes
\begin{equation}\label{NumberDensityNonExtremalPhysical}
N_{\omega} = \frac{1}{e^{2\pi(\omega-m\Omega_{h})/\ksg_{h}} - 1} ~,
\end{equation}
which represents a blackbody distribution at the Hawking temperature $T_H \equiv 
\ksg_{h}/(2\pi k_B) = \sqrt{r_s^2-4a^2}/(4\pi k_B r_s r_h)$. Clearly, the 
Hawking temperature for non-extremal Kerr black hole \cite{Iyer1979, 
Murata:2006pt, Ding:1900zz, Agullo:2010hi} depends both on its mass $M$ and the 
angular momentum parameter $a$.

\section{Extremal Kerr black holes}\label{kerr-extremal}

We note that in the \emph{extremal} limit $a\to \tfrac{1}{2} r_{s}$, the Hawking 
temperature vanishes for a non-extremal Kerr black hole. However, in this limit 
the expression of the tortoise coordinate (\ref{eq:tortoise-non-extremal}) 
becomes singular. Given the tortoise coordinate is crucial in deriving the 
Hawking effect in Kerr spacetime \cite{Liberati:2000sq, Alvarenga:2003tx, 
Angheben:2005rm, Gao:2002kz}, one is naturally led to ask whether this limit can 
be taken reliably. This provides a strong motivation to study extremal Kerr 
black hole independently in its own right. Using the definition 
(\ref{eq:diff-tortoise}) and a suitable choice of integration constant, the 
expression of the tortoise coordinate for the extremal Kerr 
black hole \emph{i.e.} with $a = r_{s}/2$, becomes
\begin{equation}\label{eq:tortoise-extremal}
 \rstar=r+r_{s} \ln 
\left(\frac{2r - r_s }{r_s}\right)-\frac{r_s^2}{2r-r_s}  ~,
\end{equation}
which differs qualitatively compared to the expression 
(\ref{eq:tortoise-non-extremal}) for non-extremal black hole.

\subsection{Relation between spatial coordinates $\xi_{-}$ and $\xi_{+}$}

In order to establish the relation between spatial coordinates $\xi_{-}$ and 
$\xi_{+}$ for extremal Kerr black hole, as earlier we consider a pivotal point 
$\xi_{-}^0$ on a $\tau_{-} = constant$ hyper-surface. A spacelike interval on 
this hyper-surface can be expressed as
\begin{equation}\label{eq:ximinus-relation}
(\xi_{-} - \xi_{-}^0)_{|\tau_{-}} = 2(\rstar^0-\rstar)_{|\tau_{-}} =
2(r^0-r)_{|\tau_{-}} \equiv \Delta ~, 
\end{equation}
where $r^0$ corresponds to $\xi_{-}^0$. On the other hand, using the 
Eqn. (\ref{eq:tortoise-extremal}), a spacelike  interval on a $\tau_{+} = 
constant$ hyper-surface, as seen by the observer $\observerplus$, 
can be expressed as
\begin{equation}\label{eq:xiplus-relation}
(\xi_{+} - \xi_{+}^0)_{|\tau_{+}} = \Delta 
+ 2r_{s} \ln \left(1 + \tfrac{\Delta}{\Delta_0}\right) - 
\tfrac{2r_{s}^2}{\Delta+\Delta_0} + \tfrac{2r_{s}^2}{\Delta_0}~,
\end{equation}
where $\Delta_0 \equiv 2(r^0 - r_{s}/2)_{|\tau_{+}}$ and again we have 
identified the interval $2(r - r^0)_{|\tau_{+}}$ as $\Delta$ using geometric 
optics approximation. By choosing $\xi_{-}^0 = \Delta_0$ and $\xi_{+}^0 = 
\xi_{-}^0 + 2r_{s} \ln (\xi_{-}^0/\sqrt{2} r_{s})-2r_{s}^2/\xi_{-}^0$, we 
can express the relation as
\begin{equation}\label{eq:near-null-relation-extremal}
 \xi_{+} = \xi_{-} + 2r_{s} \ln \left(\frac{\xi_{-}}{\sqrt{2} 
r_{s}}\right)-\frac{2 r_{s}^2}{\xi_{-}} ~.
\end{equation}
Here we note that $\xi_{+} \approx \xi_{-}$ in the region where 
$(\xi_{-}/\sqrt{2} r_{s}) \gg 1$ whereas $\xi_{+} \approx - 2 r_{s}^2/\xi_{-}$ 
for the region where 
$(\xi_{-}/\sqrt{2} r_{s}) \ll 1$. Additionally, at $(\xi_{-}/\sqrt{2} r_{s}) = 
1$, the logarithmic term $\ln \left(\xi_{-}/\sqrt{2} r_{s}\right)$ vanishes. 
Therefore, we may approximate the relation 
(\ref{eq:near-null-relation-extremal}) as
\begin{equation}\label{eq:near-null-relation-extremal-simplified}
 \xi_{+} \approx \xi_{-} -\frac{2 r_{s}^2}{\xi_{-}} ~.
\end{equation}
This approximation allows one to perform simpler analytical computations of 
the coefficient functions (\ref{FFunctionGeneral}). We may also note from the 
Eqn. (\ref{eq:near-null-relation-extremal}) that the full domain of the 
coordinate $\xi_{+}$ is $(-\infty,\infty)$ whereas it is $(0,\infty)$ for 
$\xi_{-}$ as also implied by the Eqn. 
(\ref{eq:near-null-relation-extremal-simplified}). However, as mentioned 
earlier, we shall restrict ourselves within a finite fiducial box during the 
intermediate steps.

\subsection{Evaluation of coefficient functions $F_{0}(\pm k,\kr)$}

By using the relation (\ref{eq:near-null-relation-extremal-simplified}),
the coefficient functions $F_{0}(\pm k,\kr)$ (\ref{FFunctionGeneral}) 
for an extremal Kerr black hole can be expressed as
\begin{equation}\label{F0ExtremalExpression}
 F_{0}(\pm k,\kr) = \int \frac{d\xi_{-}}{\sqrt{V_{-} V_{+}}}
\left(1 + \frac{2r_{s}^2}{\xi^{2}_{-}} \right)
 e^{i (\kr \pm k) \xi_{-} - i 2r_{s}^2 \kr /\xi_{-}} .
\end{equation}
Similar to the case of non-extremal Kerr black hole, the integral 
(\ref{F0ExtremalExpression}) is also formally divergent. Therefore, we introduce 
the standard `$i\delta$' regulation scheme with small $\delta>0$, as follows
\begin{eqnarray}\label{F0ExtremalRegulatedExpression}
 F_{0}^{\delta}(\pm k,\kr) &=& \int \frac{d\xi_{-}}{\sqrt{V_{-} V_{+}}}
\left(1 + \frac{2r_{s}^2}{\xi^{2}_{-}} \right)
~ e^{-(\delta + i)2r_{s}^2 \kr/\xi_{-}}  \nonumber \\ 
&& ~~~ \times
~~ e^{-\left[\delta |\kr \pm k| - i(\kr \pm k) \right]~\xi_{-} } ~.
\end{eqnarray}
It is easy to check that in the limit $\delta\to 0$, the regulated 
expression $F_{0}^{\delta}(\pm k,\kr)$  reduces to $F_{0}(\pm k,\kr)$. 
By introducing the variables $b_{\pm} = \sqrt{2} r_s \left[\delta|\kr \pm k| - 
i~(\kr \pm k)\right]$, $b_0 = \sqrt{2} r_s \kr (\delta 
+ i)$ and $\xi = (\xi_{-}/\sqrt{2} r_s)$, we can express the regulated 
coefficient function as
\begin{equation}\label{F0ExtremalRegulatedRedefined}
 F_{0}^{\delta}(\pm k,\kr) = \frac{\sqrt{2}~ r_s}{\sqrt{V_{-} V_{+}}} 
\int_{\xi^L}^{\xi^R} d\xi \left(1 + \xi^{-2}\right)
~ e^{-b_{\pm} \xi - b_0/\xi}  ~,
\end{equation}
where $\xi^L = (\xi_{-}^L/\sqrt{2} r_s)$ and $\xi^R = (\xi_{-}^R/\sqrt{2} r_s)$ 
are the lower and upper limits of the integration associated with the fiducial 
box. We note that there is a possibility of $(\kr - k) = 0$ \emph{i.e.} 
$b_{-}=0$, which changes the characteristics nature of the integral. Therefore, 
we evaluate this case separately.

\subsubsection{Evaluation of $F_{0}^{\delta}(-\kr,\kr)$}

For the case when $b_{-}=0$, one can evaluate the integral by defining an 
auxiliary variable $t = b_0/\xi$ as
\begin{eqnarray}\label{F0ExtremalZerobFull}
 F_{0}^{\delta}(-\kr,\kr) &=&  \frac{\sqrt{2}~ r_s}{\sqrt{V_{-} V_{+}}} 
\left[  \frac{\xi^R + b_0^{-1}}{e^{b_0/\xi^R}}   
- \frac{\xi^L + b_0^{-1}}{e^{b_0/\xi^L}}
\right.  \nonumber \\ &-& \left.
~ b_0 \Gamma\left(0,\frac{b_0}{\xi^R}\right) 
 + ~ b_0 \Gamma\left(0,\frac{b_0}{\xi^L}\right) 
\right]  ~,~
\end{eqnarray}
where $\Gamma\left(0,x\right) = \int_x^\infty dt ~t^{-1} e^{-t}$ is the 
\emph{incomplete} Gamma function. For convenience, we define the parameters 
$\gamma \equiv (V_{-}/V_{+})$ and $\mstar \equiv (\kr V_{-}/2\pi) = 
(|b_0|V_{-}/2\pi \sqrt{2} r_s)$. Using these parameters for sufficiently small 
$\xi^L$ and sufficiently large $\xi^R$ one can express the Eqn. 
(\ref{F0ExtremalZerobFull}) as
\begin{equation}\label{F0ExtremalZerobLeading}
|F_{0}^{\delta}(-\kr,\kr)|^2 =  \gamma \left[ 1 + 
\mathcal{O}\left(\frac{\ln(\mstar)}{\mstar} 
 \right) \right]  ~.
\end{equation}
Clearly, when one removes the volume regulator by taking the limit 
$\mstar\to\infty$, the coefficient function $|F_{0}^{\delta}(-\kr,\kr)|^2$ 
reduces to $\gamma$.

\subsubsection{Evaluation of $F_{0}^{\delta}(\pm k,\kr)$ with $(\kr\pm k)\ne 0$}

For the case when $b_{\pm}\ne 0$, we may define an auxiliary variable $t = 
b_{\pm} \xi$ together with $z_{\pm}^2 = 4~b_0 b_{\pm}$ to evaluate the 
coefficient functions $F_{0}^{\delta}(\pm k,\kr)$ 
(\ref{F0ExtremalRegulatedRedefined}) in terms of the \emph{modified Bessel 
functions of second kind} whose integral representations are given by 
$\mathit{K}_\nu(z) = 2^{-1} (z/2)^\nu \int_0^\infty dt ~t^{-(\nu+1)}~e^{-(t + 
z^2/4t)}$ \cite{book:olver2010nist, NIST:DLMF}. By using the identity 
$\mathit{K}_\nu(z) = \mathit{K}_{-\nu}(z)$, we can express regulated coefficient 
functions as
\begin{equation}\label{F0ExtremalNonZerobFull}
 F_{0}^{\delta}(\pm k,\kr) = \frac{\sqrt{2} ~r_s}{\sqrt{V_{-}V_{+}}} 
 \left(\frac{b_0 + b_{\pm}}{b_0~b_{\pm}}\right) 
 \left[ z_{\pm} \mathit{K}_1(z_{\pm}) \right] ~,
\end{equation}
where two boundary terms $\Delta I_L~(\sim e^{-b_0/\xi^L})$ and $\Delta I_R 
~(\sim e^{-b_{\pm} \xi^R})$ are added to complete the limits of integration. In 
the limits $\xi^L\to 0$ and $\xi^R\to \infty$ both these terms vanish. We 
may note here that the asymptotic expressions of the modified Bessel function 
are given as $\mathit{K}_1(z) \approx \tfrac{1}{z}$ for $z \ll 1$ and 
$\mathit{K}_1(z) \sim \sqrt{\tfrac{\pi}{2z}}e^{-z}$ for $z \gg 1$ 
\cite{Abramowitz1964handbook}.

\subsection{Consistency condition}

In order to satisfy the consistency condition we demand that the regulated 
coefficient functions $F_{0}^{\delta}(\pm k,\kr)$ satisfy the Eqn. 
(\ref{PoissonBracketConsistencyCondition}). For the case when $b_{\pm} \ne 0$, 
the regulated expressions of the summations can be written as
\begin{equation}\label{ExtremalSummationSpm}
\mathbb{S}_{\pm}^{\delta}(\kr) = \frac{1}{\kr~V_{-}V_{+}} 
\sum_{k>0} \frac{k}{(\kr \pm k)^2} 
\left|z_{\pm} \mathit{K}_1(z_{\pm}) \right|^2  ~.
\end{equation}
In order to carry out the summations, as in the Eqn. 
(\ref{PoissonBracketConsistencyCondition}), we may recall that $k := k_s = 2\pi 
s/V_{-}$ and $\kr := \kr_{s'} = 2\pi s'/V_{+}$ where $s$ and $s'$ are positive 
definite integers. Therefore, we can express the \emph{lhs} of Eqn. 
(\ref{PoissonBracketConsistencyCondition}) for extremal Kerr black hole as
\begin{eqnarray}\label{ExtremalSmSpDifference}
\mathbb{S}^{\delta}_{-}(\kr) - \mathbb{S}^{\delta}_{+}(\kr) =
|F_{0}^{\delta}(-\kr,\kr)|^2 + \frac{\gamma\zeta(2)}{2\pi^2} +
\frac{\gamma~\mathcal{S}(1,\infty)}{4\pi^2 \mstar} ~.~~
\end{eqnarray}
Here the auxiliary summation function $\mathcal{S}(s_0,s_1)$ is introduced as
\begin{eqnarray}\label{Ss0s1Definition}
\mathcal{S}(s_0,s_1) &=& \sum_{s=s_0}^{s_1}  \left[
\frac{2\mstar}{s^2}  \left\{ |\tilde{z} \mathit{K}_1(\tilde{z})|^2 - 1 
\right\}  \right. \nonumber \\
&+& \left. \frac{(s-\mstar)}{s^2}  \left\{ |\tilde{z} 
\mathit{K}_1(\tilde{z})|^2 - |\tilde{z} \mathit{K}_1(|\tilde{z}|)|^2 \right\} 
\right] ,~~~
\end{eqnarray}
where $\tilde{z} \equiv \tilde{z}(s) = \sqrt{4|b_0|^2s/\mstar}~ (\delta+i)$. In 
order to arrive at the Eqn. (\ref{ExtremalSmSpDifference}) we have included the 
possibility of $(\kr-k)=0$ which in turn demands that $\gamma$ must be a 
ratio of two positive definite integers \emph{i.e.} $\gamma$ must be a 
\emph{rational} number.

Given the function $\mathit{K}_1(z)$ satisfies $\lim_{z\to 0} 
\left|z \mathit{K}_1(z)\right| = 1$, $\lim_{z\to\infty} 
\left|z \mathit{K}_1(z)\right| = 0$ and it has no other pole, there 
exist upper bounds $d_1$ and $d_2$ such that $\left|\tilde{z} 
\mathit{K}_1(\tilde{z})\right|^2 \le d_1$ and $\left|\tilde{z} 
\mathit{K}_1(|\tilde{z}|)\right|^2 \le d_2$ for all allowed values of 
$\tilde{z}$. For a given value of $|b_0|$ the removal of volume regulator 
$V_{-}\to \infty$ is achieved by taking the limit $\mstar \to \infty$. In such 
limit we may choose two numbers $\lambda_1$ and $\lambda_2$ such that 
$|\tilde{z}(\lambda_1 \mstar)| = 2|b_0| \sqrt{\lambda_1} \ll 1$ and 
$|\tilde{z}(\lambda_2 \mstar)| \gg 1$. We note that for a given $|b_0|$ and  
sufficiently large $\mstar$, both $\lambda_1 \mstar \gg 1$ and $\lambda_2 \mstar 
\gg 1$. Consequently, we may express the summation as
\begin{equation}\label{ExtremalSplusAuxiliary}
\mathcal{S}(1,\infty) = \mathcal{S}(\lambda_1\mstar,\lambda_2\mstar) + 
\mathcal{S}(\lambda_2\mstar+1,\infty) ~.
\end{equation}
where we have used $\mathcal{S}(1,\lambda_1\mstar-1) = 0$ as the corresponding
$|\tilde{z}| \ll 1$. Given the form of $\mathcal{S}(s_0,s_1)$, we can 
approximate the summation by an integration for large $s$. Thereafter, we can 
establish an inequality
\begin{equation}\label{ExtremalSpmAux1}
\mathcal{S}(\lambda_1\mstar,\lambda_2\mstar) \le
(d_1 + d_2)\left[ \ln \left(\frac{\lambda_2}{\lambda_1}\right) + 
\frac{\lambda_2 - \lambda_1}{\lambda_2 \lambda_1} \right] ~.
\end{equation}
Similarly, the asymptotic form of $\mathit{K}_1(\tilde{z})$ leads to
\begin{equation}\label{ExtremalSpmAux2}
\mathcal{S}(\lambda_2\mstar+1,\infty) = \frac{\pi}{2\delta} 
e^{-4|b_0|\delta\sqrt{\lambda_2}} \left[ 1 + \mathcal{O}(\delta) \right] ~.
\end{equation}
We note the critical role that is played by the integral regulator $\delta$ in 
the Eqn. (\ref{ExtremalSpmAux2}). In particular, in 
the absence of the integral regulator $\delta$ the summation would have 
diverged. In the limit $\mstar\to\infty$. \emph{i.e.} when the volume 
regulators are removed for a fixed $\delta$, we can express the Eqn. 
(\ref{ExtremalSmSpDifference}) 
as
\begin{equation}\label{ExtremalSpmRelationLimit}
\mathbb{S}_{-}(\kr) - \mathbb{S}_{+}(\kr) = \gamma \left[1 +  
\frac{1}{2\pi^2} ~\zeta(2)   \right] ~.
\end{equation}
Using the value of the \emph{Riemann zeta function} $\zeta(2) = \tfrac{1}{6} 
\pi^2$, we conclude that in order to satisfy the required consistency 
condition one must demand $\gamma = (12/13)$ which is indeed a \emph{rational} 
number as required. Together with the Eqn. 
(\ref{eq:near-null-relation-extremal-simplified}) we may express the 
consistency condition also as
\begin{equation}\label{ExtremalConsistencyConditionXipXim}
\left( \frac{\xi_{-}^L}{\sqrt{2} r_s} \right) = 12 
\left( \frac{\sqrt{2} r_s} {\xi_{-}^R} \right) ~.
\end{equation}
Clearly, the requirement that Poisson brackets of both observers be 
simultaneously satisfied also for extremal Kerr black holes, demands that the 
volume regulators $\xi_{-}^L$ and $\xi_{-}^R$ are not to be treated 
independently but should be varied together as given in the Eqn. 
(\ref{ExtremalConsistencyConditionXipXim}). We may also point out that when 
volume regulators are removed then the integral regulator $\delta$ fully drops 
off from the expression.

\subsection{On inconsistency of Bogoliubov coefficients}

We would like to note that as reported in \cite{Alvarenga:2003tx, 
Liberati:2000sq, Alvarenga:2003jd}, the Bogoliubov coefficients for extremal 
black holes fail to satisfy the analogous consistency condition 
$\int d\omega'd\omega'' (\alpha_{\omega\omega'} 
\alpha_{\omega''\omega'}^{*} -\beta_{\omega\omega'} 
\beta_{\omega''\omega'}^{*})=1$. The key reason behind this failure of the 
Bogoliubov coefficients lies in the improper approximation made in the relation 
between the null coordinates, given by $u = C/(v_0-v)$ \cite{Alvarenga:2003tx}. 
This relation is used in evaluation of the Bogoliubov coefficients and is 
analogous to the Eqn. (\ref{eq:near-null-relation-extremal-simplified}) between 
the near-null coordinates here. It may be emphasized that one would encounter 
the same failure in satisfying the consistency condition even here, had one used 
the approximation $\xi_{+} \approx - (2 r_{s}^2/\xi_{-})$ instead of the Eqn. 
(\ref{eq:near-null-relation-extremal-simplified}). Firstly, this approximation 
would fail to fully cover the domain of $\xi_{+}$, given the domain of $\xi_{-}$ 
is $(0,\infty)$. This is unlike the analogous approximation for non-extremal 
Kerr black hole  $\xi_{+}  \approx \ln (\ksg_{h}\xi_{-})/\ksg_{h}$ 
(\ref{eq:final-near-null-relation}) which covers the full domain of $\xi_{+}$. 
Secondly, this approximation would have lead the expression $b_{\pm}$ to be 
$\sim (\pm k)$ rather than $\sim (\kr\pm k)$. Due to this one would have missed 
the possibility of $(\kr-k) = 0$ which directly gives rise to the leading term 
$|F_{0}^{\delta}(-\kr,\kr)|^2$ in the consistency condition 
(\ref{ExtremalSmSpDifference}). Furthermore, even the term $\zeta(2)$ in the 
same consistency condition originates because $|b_{-}|^2 \ne |b_{+}|^2$. Without 
these terms being present even here one would have failed to satisfy the 
required consistency condition. We may add that for non-extremal Kerr black 
hole even if one considers the relation to be  $\xi_{+} \approx \xi_{-} + \ln 
(\ksg_{h}\xi_{-})/\ksg_{h}$, the conclusion there remains unaffected.

\subsection{Number density of Hawking quanta}

We have shown that the expectation value of the number density operator 
corresponding to the Hawking quanta in Fock quantization can be expressed
in terms of $\mathbb{S}_{+}(\kr)$ (\ref{NumberDensityEVGeneral}). For 
convenience, 
we define the following auxiliary summation 
\begin{equation}\label{ExtremalSplusAuxiliary}
\mathcal{S}_{+}(s_0,s_1) = \sum_{s=s_0}^{s_1} \frac{(s-\mstar)}{s^2} 
\left|\tilde{z} \mathit{K}_1(|\tilde{z}|)\right|^2  ~.
\end{equation}
The regulated expression of the summation $\mathbb{S}_{+}(\kr)$ can then be 
written as 
\begin{equation}\label{ExtremalSplusRegulated}
\mathbb{S}^{\delta}_{+}(\kr) = \frac{\gamma}{4\pi^2 \mstar} \left[ 
\mathcal{S}_{+}(\mstar,\lambda_2\mstar) + 
\mathcal{S}_{+}(\lambda_2\mstar+1,\infty) \right] ~.
\end{equation}
As earlier, for large $\mstar$ we can approximate the summation by an 
integration to establish an inequality as
\begin{equation}\label{ExtremalSplusAux1}
\mathcal{S}_{+}(\mstar,\lambda_2\mstar) 
\le d_2 \left[ \ln \left(\lambda_2\right) - 1 
+ \frac{1}{\lambda_2} \right]  ~.
\end{equation}
Similarly, by using the asymptotic form of $\mathit{K}_1(\tilde{z})$, we can 
evaluate
\begin{equation}\label{ExtremalSplusAux2}
\mathcal{S}_{+}(\lambda_2\mstar+1,\infty) = \frac{\pi}{2} 
e^{-4|b_0|\sqrt{\lambda_2}} \left[ 1 + 
\mathcal{O}\left(\frac{1}{\lambda_2}\right) \right] ~.
\end{equation}
We note from  the Eqn. (\ref{ExtremalSplusAux1}) and (\ref{ExtremalSplusAux2}) 
that their leading terms are independent of $\mstar$. Therefore, in the limit 
${\mstar\to\infty}$, the Eqn. (\ref{ExtremalSplusRegulated}) implies that 
$\mathbb{S}^{\delta}_{+}(\kr) \le 0$. On the other hand, by definition
$\mathbb{S}^{\delta}_{+}(\kr) \ge 0$ and hence
\begin{equation}\label{ExtremalSplusLimit}
\lim_{\mstar\to\infty} \mathbb{S}^{\delta}_{+}(\kr) = 0  ~.
\end{equation}
We note that when volume regulators are removed then the integral 
regulator $\delta$ also drops off fully. Therefore, the expectation value of the 
number density operator (\ref{NumberDensityEVGeneral}) associated with the 
Hawking quanta of physical frequency $\omega$ for extremal Kerr black hole is 
given by
\begin{equation}\label{NumberDensityEVExtremal} 
N_{\omegat} = N_{\omega-m\Omega_{h}} = \langle\hat{N}_{\kr}^{+}\rangle = 0 ~,
\end{equation}
where $\omegat = \kr > 0$. In other words, the extremal Kerr black hole does not 
emit Hawking radiation.

\section{Discussion}\label{discussion}

In summary, we have shown here that one can perform an exact derivation of the 
Hawking effect using Hamiltonian based canonical formulation for both 
non-extremal and extremal Kerr black holes. In order to do so we have extended 
the scope of the so-called near-null coordinates which were recently introduced 
for canonical derivation of Hawking effect in Schwarzschild spacetime 
\cite{Barman:2017fzh}. In the context of extremal Kerr black holes it is 
usually believed that extremal black holes do not emit Hawking radiation as one 
would conclude by taking the extremal limits of non-extremal black holes. 
However, whether one can make such conclusion starting from an extremal black 
hole is debated in the literature \cite{Vanzo:1995bh,Liberati:2000sq, 
Alvarenga:2003jd, Alvarenga:2003tx}. These debates stem from the fact that the 
associated Bogoliubov coefficients that relate the ingoing and the outgoing 
field modes do not satisfy the required consistency condition. Therefore, these 
Bogoliubov coefficients are not considered to be reliable for extremal black 
holes. In the canonical formulation the analogous consistency condition arises 
from the requirement of the Poisson bracket of field modes and their conjugate 
momenta be simultaneously satisfied for different observers. Here we have shown 
that in the canonical derivation the required consistency condition is satisfied 
also for extremal Kerr black holes. We have also pointed out the reason behind 
the reported failure of Bogoliubov coefficients to satisfy the required 
condition. Further, we have shown that the expectation value of the associated 
number density operator vanishes for the extremal Kerr black holes. This aspect 
reaffirms that the extremal Kerr black holes do not emit Hawking radiation.

The canonical derivation of the Hawking effect for Kerr black holes as presented 
here provides an initial stage for the study of Hawking effect in the context of 
the so called polymer quantization \cite{Ashtekar:2002sn,Halvorson-2004-35}, 
specially as applied in \cite{Hossain:2014fma,Hossain:2016klt, 
Hossain:2015xqa,Barman:2017vqx}. Additionally, the method as developed for Kerr 
spacetime can be generalized for other similar spacetimes such as 
Reissner-Nordstr\"om and Kerr-Newman \cite{Fabbri:2000xh, Dehghani:2010zzb, 
Kamali:2016qyj, Liu:2007zza, Jiang:2006ha, Wu:2001ib, Zhang:2005uh, 
Vieira:2014waa, Iso:2006ut, Umetsu:2009HRKN, Lin:2009wm} in a straightforward 
manner.

\begin{acknowledgments}
We thank Gopal Sardar and Chiranjeeb Singha for discussions. S.B. would like to 
thank IISER Kolkata for supporting this work through a doctoral fellowship. 
\end{acknowledgments}

\end{document}